\documentstyle[multicol,aps,prl]{revtex}

\begin{document}
\widetext

\draft
\title{Spatial and temporal turbulent velocity and vorticity power spectra from sound scattering.}
\author{Shahar Seifer$^1$ and Victor Steinberg$^{1,2}$}

\address{$^1$Department of Physics of Complex Systems, \\
Weizmann Institute of Science, Rehovot, 76100 Israel, and \\
$^2$International Center for Theoretical Physics, Strada Costiera
11, I-34100, Trieste, Italy}

\date{\today}
\maketitle
\begin{abstract}
By performing sound scattering measurements with a detector array
consisting of 62 elements in a flow between two counter-rotating
disks we obtain the energy and vorticity  power spectra directly
in both spatial and temporal domains. Fast accumulated statistics
and large signal-to-noise ratio allow to get high quality data
rather effectively and to test scaling laws in details.
\end{abstract}


\begin{multicols}{2}

\narrowtext

One of the challenging experimental tasks in studies of turbulent
flows is developing new tools to measure  spectral characteristics
of velocity and vorticity fields in a spatial domain. Such
measurements will allow direct comparison of  experimental data
with a theory without exploiting the Taylor hypothesis
particularly in those cases when its use is rather questionable.
Currently, from conventional methods in use only particle image
velocimetry (PIV) is capable of performing this task on the
expense of a very heavy data analysis to get representative
statistics and a reasonable signal-to-noise ratio.

In this Letter we present results on direct measurements of the
velocity and vorticity fluctuations spatial and temporal spectra
by using sound scattering amplitude measurements. We achieve this
goal by simultaneous acquisition of sound pulses on 62 sound
detectors, arranged in front of a linear emitter in the same plane, as has been already described in Ref.\cite{seifer}.
There were several attempts in the past to experimentally probe
turbulent flow characteristics by the sound scattering method.
These experiments were mostly concentrated on studies of vorticity
distribution \cite{oljaca} as well as temporal dynamics of
vorticity fluctuations\cite{fauve1,baudet} and a large scale
circulation in high Reynolds numbers flows\cite{fauve2,seifer}.

As we pointed out\cite{seifer} the main obstacles to get reliable
and complete information about the velocity field from sound
scattering are a finite width of a sound beam and far field
approximation required by the
theory\cite{kraichnan,fabricant,lund}. Since an integral scale of
the velocity fluctuations is of the order of the beam width, the
former limitation is released for the turbulent velocity
fluctuations. Thus, to get spatial information about the velocity
spectra with sufficient dynamical range from the sound scattering
data one needs to exploit a large number of elements in the sound
detector array, to use lock-in detection technique to improve
signal-to-noise ratio, and to apply the Huygens far-field
construction method\cite{seifer}. Then the amplitude of the
complex wave function of the scattered sound in the far-field
limit can be related to the Fourier transform of the velocity
field (and also the vorticity) in a scattering plane as follows
\cite{fabricant,lund}:

$$\Psi_{scat}(\vec{r},t)=\frac{1}{c}\frac{(2\pi
k_0)^2\exp{(i\pi/4)}}{\sqrt{2\pi
k_0r}}\cos{\theta}F_{\vec{k_s}}\{v_x\}, \eqno(1)$$

where $\Psi_{scat}=\Psi-\Psi_{rest}$, $\Psi$ and $\Psi_{rest}$ are
the complex wave functions that describe the sound pressure
oscillations in the presence of a flow and without it,
respectively; $k_0$ and $c$ are the sound wave number and the
velocity, respectively; $\theta$ and $k_s$ are the scattering
angle and the wave number, respectively, that are related to each
other via $k_s\equiv |\vec{k_s}|=2k_0\sin{\theta/2}$; and
$r=|\vec{r}|$ is the distance from the center of a scattering
region till a detector. $F_{\vec{k_s}}\{v_x\}$ is the 2D Fourier
transform of the velocity component in the forward direction of the beam that is related to the Fourier transform
of the vorticity via
$F_{\vec{k_s}}\{v_x\}=\frac{i}{2k_0}\cot{(\theta/2)}F_{\vec{k_s}}\{(\nabla\times\vec{v})_z\}$.

Since in many
cases there exist severe technical obstacles to obtain a signal in
a far-field limit, a construction of the far-field scattering wave
function from the near-field measurements was suggested and used
\cite{seifer}. The method is based on a mathematical description
of the Huygens principle in optics and derived from the
Rayleigh-Sommerfeld integral\cite{mar} as:
$$\Psi(r_f,y)^{ff}_{scat}=\int\frac{k_0 i^{3/2}dy'}{\sqrt{2\pi
k_0(r_f-r_d)}}$$
$$\exp{(\frac{ik_0(y-y')^2}{2(r_f-r_d)})}\Psi(r_d,y')^d_{scat},
\eqno(2)$$

where $r_d$ and $r_f$ are the distances measured from the cell
center till the detector and the far-field region, respectively;
$\Psi(0,y')^{d}_{scat}$ and $\Psi(x,y)^{ff}_{scat}$ are the
scattering wave functions at the detector (as measured) and at the
far-field, respectively\cite{seifer}.

The experimental set-up is described in details in
Ref.\cite{seifer}. The von Karman swirling water flow was produced
between two counter-rotating disks of a diameter $2R=280$ mm with
four triangular blades of 20 mm high and 5 mm thick and with rims.
They are driven by two dc brushless motors, which velocity is
controlled with a stability of about 0.1\% via optical encoders.
This set-up is well recognized to generate a strong intensity
turbulent flow in a confined region (see for example
\cite{fauve1}). The flow is confined by a perspex cylinder of an
inner diameter 290 mm and 320 mm in height and disks separation
205 mm. By changing a rotation frequency the Reynolds number,
$Re=2\Omega R^2/\nu$, is varied between $2.5\cdot 10^5$ and
$1.7\cdot 10^6$ that corresponds to the Taylor microscale Reynolds
number $R_{\lambda}$ between 200 and 570. Here $\Omega$ is the
angular velocity of the disk, $\nu$ is the kinematic viscosity,
and the energy dissipation, $\epsilon=4.9\cdot 10^{-18}Re^3$ W/kg,
and the rms of the velocity fluctuations in the middle plane,
$V_{rms}=0.5\cdot 10^{-6}Re$ m/sec, obtained from the global
torque and the hot wire anemometry (HWA) measurements,
respectively\cite{shahar}.

The sound
scattering measurements were conducted in the middle plane between
the disks and in the plane at 30 mm below it. The emitter was 100
mm long and 10 mm wide, and the size of the scattering region was
defined by the width of the detector designed as a linear array of
62 acoustic detectors with 1 mm spacing and $62\times 10$ mm
active area (from Blatek). A range of frequencies covered in the
experiment was between 1 and 7 Mhz. The acquisition system is
built in a heterodyne scheme that is based on 62 lock-in
amplifiers with 62 preamplifiers. The details of the design and
its operation are presented in Ref.\cite{seifer}. A typical sound
propagation time through the cell is about 200 $\mu$sec that is a
typical freezing time segment. Within this period one pulse is
sent, and the flow is almost frozen. Since an each sound pulse has
a sufficiently high signal-to-noise ratio, it is used to construct
a sound far-field wave function from sound scattering signals
acquired by all detectors. Obtained as a function of time the wave
function provides a possibility to study statistics of velocity
(vorticity) fluctuations in a turbulent flow.

A scattering wave
 function emitted from a single transducer and obtained by the detector array
 provides information about the velocity (vorticity) field only
 on a single curve in a (($k_{s})_{x},(k_{s})_{y}$) plane (see Fig.1, left inset, solid line).
 To get information on other curves in the plane (see Fig.1, left inset, dashed lines)
 requires to use sound beams emitted by different transducers in many different directions
 simultaneously. Then complete structure functions of the
 velocity (vorticity) fluctuations can be retrieved without any assumption about isotropy
 and homogeneity of a turbulent flow. However, with only one emitter and the
 detector array we should rely on the isotropy and homogeneity assumptions of
 the turbulent flow under studies. In this case the sound
 scattering provides direct measurements of the energy
 spectrum as well as the Fourier transform of the vorticity structure function in a spatial
 domain that are related as $E(k_s)=\frac{6\pi^3}{Ak_s}
|F_{\vec{k_s}}\{(\nabla\times \vec{v})_z\}|^2.$ Here the kinetic
energy per unit mass is defined as
$\int{E(k)dk}=\frac{3}{2A}\int{v_x^2d^2r}$, where
 $A$ is the cross-section area of a sound
beam. We neglect
 velocity variations in the direction perpendicular to the beam
propagation by averaging over the beam thickness. In Fig.1 we
present a typical result on the time-averaged velocity Fourier
transform in the far-field obtained from the scattering wave
function via Eq.(1) and the far-field reconstruction according to
Eq.(2). The same function but observed directly in the detector
plane looks drastically different (see right inset in Fig.1).
These data are the result of averaging on 60,000 sound pulses at
1.8 kHz repetition rate and at frequency 3MHz taken in the von
Karman swirling flow at $Re=1.5\cdot 10^6$.

In Fig.2 the resulting energy spectrum as a function of the wave
number is shown. The dotted line denotes the "-5/3" slope
according to the Kolmogorov law to demonstrate that indeed in some
range of the wave numbers the spectrum follows the expected law.
It can be compared with the results on the energy spectra obtained
by PIV, where though a shorter scaling region is also
observed\cite{shahar}.  In the sound measurements shown in Fig.2
the wave numbers are limited by the range of values $0.1<k<0.8$
mm$^{-1}$, outside of which the spectrum cannot be retrieved. The
lower side of the range of $k$ is limited by the beam width, i.e.
one cannot get information on a scale exceeding the beam width (in
the energy spectra from PIV the scaling region begins already at
$k=0.06$ mm$^{-1}$ for $Re>10^{6}$) . On the higher side of the
range the ultimate limit is determined by the size of the element
in the detector array, namely $k_{max}\leq 2\pi$ mm$^{-1}$
corresponding to 1 mm detector array spacing. However, even before
this limit is reached, the energy spectrum is cut on the higher
 wave number side by the visibility at large scattering angles through the detector
aperture. It means that some sound rays are blocked by the limited
length of the detector array. According to the detector array
length and the cell diameter at angles exceeding about $6^{\circ}$
the visibility starts to deteriorate (see the inset in Fig.2). The
corresponding limit is
$k_{s_{(lim)}}=2k_0\sin{(\theta_{max}/2)}\simeq 0.1k_0$. It gives
about 1.3 mm$^{-1}$ at 3MHz compared to $0.8$ m$^{-1}$ observed.
We tested this relation at various frequencies, and the results on
the energy spectra taken at different sound frequencies are
presented in Fig.3. One finds that by changing the frequency from
1 up to 7MHz the upper wave number limit moves linearly toward the
highest value of about 1.5 mm$^{-1}$, but still far away from
$k_{max}=2\pi$ mm$^{-1}$.

    Useful information in terms of the turbulent flow energy dissipation,
$\epsilon$, can be gained from combined presentation of the energy
spectra obtained from the sound scattering measurements at
different Reynolds numbers. There is a known scaling law for the
energy spectra \cite{nelkin,sad,frisch} that appears as the result
of plotting the scaled energy density spectrum
$\epsilon^{-2/3}\eta^{-5/3}E(k)$ versus $k\eta$, where $\eta$ is
the Kolmogorov dissipation scale defined as
$\eta=(\nu^3/\epsilon)^{1/4}$\cite{frisch}. The idea is to find
the best match between the scaled spectra at different $Re$ with
fitting values of $\epsilon$. It turns out that the scaling exists
for the data at all values of $Re$, and the results for the sound
scattering in the middle plane (curve 1) and in the plane at 30 mm
below it (curve 2) are shown in Fig.4. Each data set for each $Re$ consists
of $4\cdot 10^6$ points. The dependence of the
energy dissipation, $\epsilon$, on $Re$ is found to be
$\epsilon\sim Re^{3.1\pm 0.1}$ with a good quality of the fits.
The exponent value is rather close to 3, the expected one
according to the dimensional analysis \cite{shahar,cadot}(compare with the expression
for $\epsilon$ presented above). The Kolmogorov
constant $C$ in the Kolmogorov equation
$\epsilon^{-2/3}\eta^{-5/3}E(k)=C(k\eta)^{-5/3}$ is determined
experimentally from the fit of the plots in Fig.4, and the value
is $C\simeq 0.8$ for the curve (1) and $C\simeq 0.9$ for the curve
(2).

We also calculate the integral scale in the flow at $Re=1.2\cdot
10^6$, using\cite{leseiur}
$L_{int}=(3\pi/4)\int{k^{-1}E(k)dk}/\int{E(k)dk}=40\pm 10$ mm,
based on the system scale 0.02 mm$^{-1}$ and the scaling region
$0.06<k<1.5$ mm$^{-1}$. This value occurs to be rather close to
the beam width, and the corresponding wave number is located close
to the lower end of the wave number range of the spectrum.

To test the Taylor hypothesis\cite{frisch,leseiur} for the
swirling flow the energy spectrum in the frequency domain is
calculated as $E(f)=\frac{1}{T}\int_0^T{E(t)\exp({-i2\pi ft})dt}$
and
$E(t)=\int{\frac{6\pi^3}{Ak_s}|F_{\vec{k_s}}{(\nabla\times\vec{v(t)})_z}|^2d^2k_s}$.
A proper energy spectrum can be obtained only, if the lowest $k_s$
are available due to the pole at $k_s^{-1}$ in the integrand. So
we turn out to the Fourier transform of the vertical vorticity. In
Fig.5 (upper curve) we present the time-averaged power spectrum of
the vorticity as a function of $k$ (based on $2\cdot 10^7$ points)
at $Re=1.2\cdot 10^6$ obtained at the sound frequencies 2.5 MHz
(asterisks) and 5.8MHz (diamonds). The dashed line with the
scaling exponent "-2/3"
 represents the expected dependence according to the Kolmogorov
predictions\cite{frisch,leseiur}. It can be compared with the
space-averaged over the beam area Fourier transform of the
enstrophy in the frequency domain
$N(f)=\frac{1}{T}\int_0^T{N(t)\exp{(-i2\pi ft)}dt}$, where
$N(t)=\frac{1}{A}\int_A{|(\nabla\times\vec{v(t)})_z|^2d^2r}=(2\pi)^2
\int{|F_{\vec{k_s}}{(\nabla\times\vec{v(t)})}|^2d^2k_s}$. The
latter shows a rather wide scaling region (see the lower curve in
Fig.5). The advection velocity, $V_T$, found from the two plots in
Fig.5 is $V_T=2\pi f/k\simeq 0.6$ m/sec that is rather close to
the average velocity in this experiment\cite{shahar}.

We would like to point out also that the mean enstrophy value,
marked on Fig.5 by asterisk on the left hand-side ordinate axis,
is about an order of magnitude larger than one corresponding to a
rigid body rotation with the same rotation velocity of 300 rpm and
the vorticity of 63 sec$^{-1}$, or the enstrohy of about 4000
sec$^{-2}$. In the co-rotational disks geometry at the same
parameters and with a large scale single vortex flow configuration
the enstrophy is also much lower\cite{seifer}.

  This work is partially supported by Israel Science Foundation
grant, by Binational US-Israel Foundation grant, and by the
Minerva Center for Nonlinear Physics of Complex Systems.

\begin{figure}
\caption {Time averaged velocity structure function obtained by
sound scattering
 via the far-field construction taken at $Re=1.5\cdot 10^6$ and 3 MHz.
 The right inset: the same property obtained from sound
scattering data at the detector without far-field construction.
The left inset: path in the wave number plane (solid line),on
which information on the velocity (vorticity) spectrum for a given
beam direction is obtained; other curves (dashed lines) contain
information on scattering from sound beams emitted in various
directions.}
 \label{figa}
 \end{figure}

\begin{figure}
\caption { Energy spectrum derived from the data presented in
Fig.1. The inset: scheme of the aperture limit for the sound
detector array.}

\label{figb}
 \end{figure}

\begin{figure}
\caption {(color online) Energy spectra taken at various sound
frequencies from 1 up to 7 MHz at $Re=1.2\cdot 10^6$. }

 \label{figc}
 \end{figure}

\begin{figure}
\caption { (color online) Scaled energy spectra as a function of
reduced wave number at various $Re$: (1) at $h=0$ (2) at $h=-30$
mm. The dashed lines show the Kolmogorov scaling law with the
exponent "-5/3".}

\label{figd}
 \end{figure}

\begin{figure}
\caption {Upper curve: Power spectra of time averaged vorticity as
a function of $k$ taken at $Re=1.2\cdot 10^6$ and frequencies 2.5
(asterisks) and 5.8 (diamonds) MHz. Lower curve: Power spectra of
space-averaged enstrophy as a function of $f$. The dashed lines
show the Kolmogorov scaling with the exponent "-2/3".}

\label{fige}
 \end{figure}

\end{multicols}
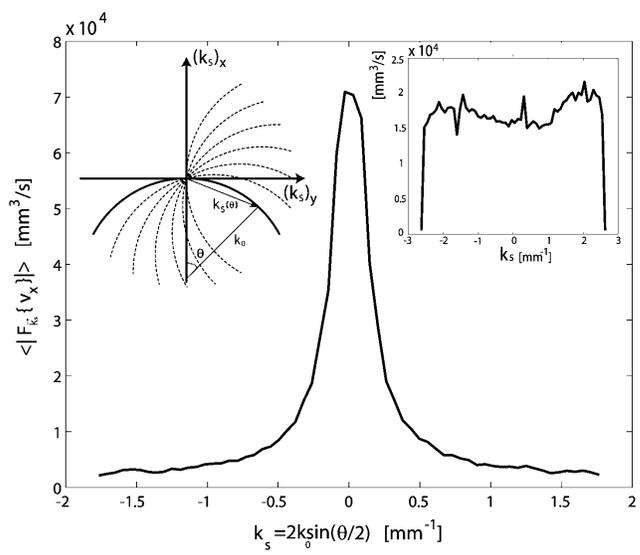
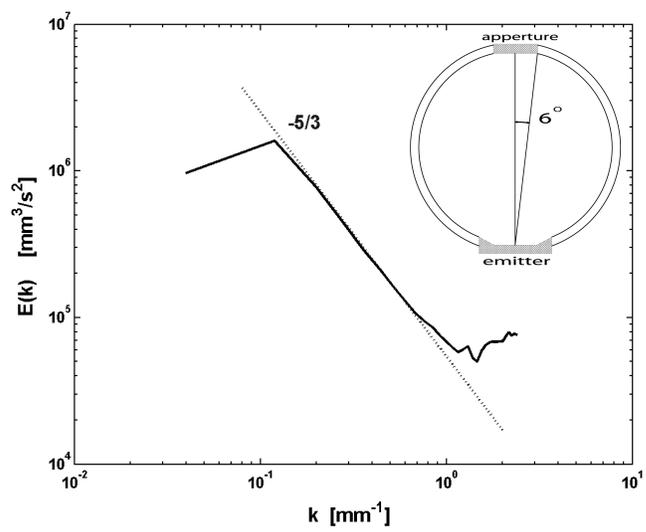
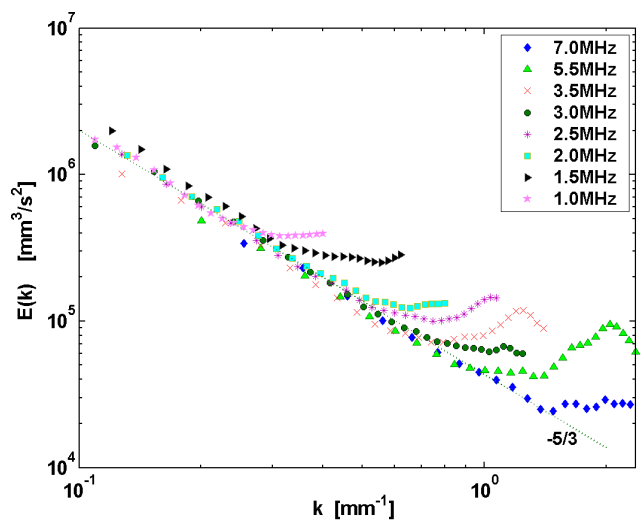
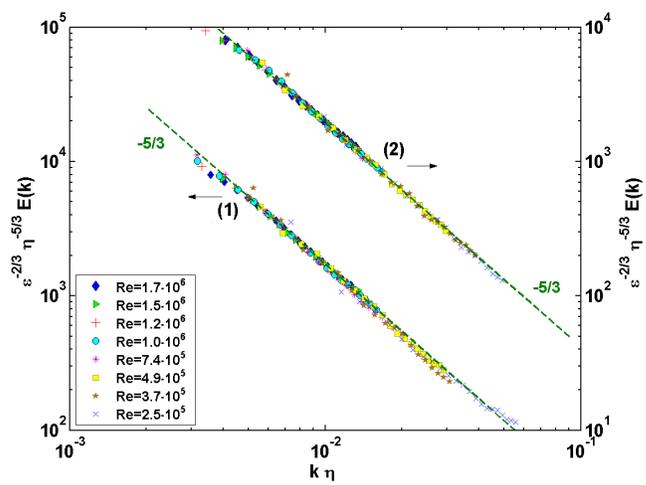
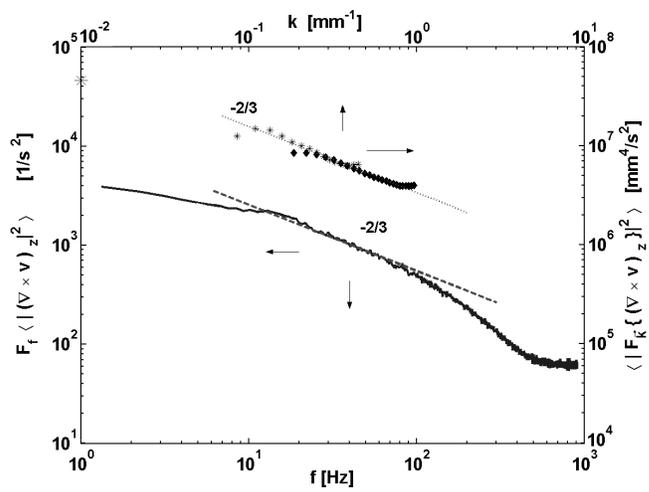
\end{document}